\documentclass[manuscript]{aastex6}

\usepackage{multirow}

\newcommand{\cnrt}{CN7820}
\newcommand{\cnro}{CN7065}
\newcommand{\ctwo}{$\rm C_2$}
\newcommand{\cn}{C-N}
\newcommand{\ch}{C-H}
\newcommand{\CR}{C-R}
\newcommand{\cj}{C-J}
\newcommand{\K}{$K_s$}
\newcommand{\JK}{$J-K_s$}
\newcommand{\JH}{$J-H$}

\shorttitle{LAMOST carbon stars}
\shortauthors{Ji et al.}

\begin{document}


\title{Carbon stars from LAMOST DR2 data}


\author{Wei Ji\altaffilmark{1,2}, Wenyuan Cui\altaffilmark{1,3}, Chao Liu\altaffilmark{2},
Ali Luo\altaffilmark{2}, Gang Zhao\altaffilmark{2}, Bo Zhang\altaffilmark{1}}
\affil{$^1$Department of Physics, Hebei Normal University, Shijiazhuang 050024, China;
wenyuancui@126.com}
\affil{$^2$Key Lab of Astronomy, National Astronomical Observatories, Chinese Academy of Sciences, Beijing 100012, China;
liuchao@bao.ac.cn}
\affil{$^3$School of Space Science and Physics, Shandong University at Weihai, Weihai 264209, China}


\begin{abstract}

In this work, we present the new catalog of carbon stars from the LAMOST DR2 catalog. 
In total, 894 carbon stars are identified from multiple line indices measured from the stellar spectra. 
Combining the CN bands in the red end with \ctwo\ and other lines, we are able to identify 
the carbon stars. Moreover, we also classify the carbon stars into spectral
sub-types of \ch, \CR, and \cn. These sub-types approximately show distinct features in the multi-dimensional line
indices, implying that in the future we can use them to identify carbon stars from larger
spectroscopic datasets. Meanwhile, from the line indices space, while the \cn\ stars are clearly separated from the
others, we find no clear
separation between \CR\ and \ch\ sub-types. The \CR\ and \ch\ stars seem to smoothly transition from one to another. This may hint
that the \CR\ and \ch\ stars may not be different in their origins but look different in their 
spectra because of different metallicity. Due to the relatively low spectral resolution 
and lower signal-to-noise ratio, the ratio of $^{12}$C/$^{13}$C is not measured and thus the \cj\ stars are not identified.

\end{abstract}



\section{Introduction} \label{sec:intro}

Carbon stars, recognized by \citet{sec69}, are peculiar objects with their optical spectra characterized
by strong carbon molecular bands, namely CH, CN, and \ctwo, as well as SiC$_2$ and C$_3$ in the cooler stars. 
Compared to normal stars, carbon stars show an 
inversion of the C/O ratio (C/O\,$>1$). Most carbon stars in the galaxy are dwarf carbon stars.
Both dwarf carbon stars and \ch\ stars inherited their atmospheric carbon by mass transfer
from an asymptotic giant branch (AGB) companion which is now a white dwarf. Most such white
dwarfs have already cooled sufficiently to be undetectable in
spectroscopy. However, the inversion of the C/O ratio in the cool and luminous 
carbon stars (C-N type) is due to the currently undergoing third dredge-up in the AGB phase. 
The origin of the enrichment of \CR\ stars is still in debate, 
since long-term radial-velocity observations revealed that they seem not to be in binary systems \citep{mcc97}, which 
leads to the conclusion that they are the product of a merger and mergers are known to produce fast rotators. 
Theoretical works suggested that the merging of a He white dwarf with a red giant branch star is the most favorable progenitor,
because a strong off-center He-flash led by rotation of the He-core formed after the merger is expected to  
provoke the mixing of carbon into the envelope \citep{izz07,dom10}.
Almost all the excess of carbon in carbon stars is 
produced through the triple-$\alpha$ reaction during helium fusion in red giant stars.

As the carbon-rich AGB stars are very luminous and easily identified by their strong band heads in their spectra, 
they are usually used as kinematical and dynamical probes of the Galaxy at large distance. 
For instance, carbon stars have been used to measure the velocity dispersions of the 
Galactic halo \citep[e.g.,][]{mouse5,bot91} and the Galactic rotation curve \citep{dem07,dem09,bat3}. 
The mean apparent magnitude of carbon stars has also been used to estimate the distance moduli 
of NGC\,205 and NGC\,300 \citep{ric84,ric85}.

Up to now, a series of surveys have been conducted for searching the Galactic carbon stars such as 
the Automatic Plate Measuring (APM) survey \citep{tot98,iba01}, the First Byurakan Spectral Sky 
Survey \citep{gig98} and infrared objective-prism surveys \citep[][and references therein]{alk01}. 
In particular, \citet{alk01} published a catalogue containing 6\,891 carbon stars, which is the revised version of the 
General Catalogue of Cool Carbon Stars maintained by B. C. Stephenson of the Warner and 
Swasey Observatory \citep{ste73,ste89}. 

Carbon stars are mainly identified from the optical C$_2$ Swan bands, the near-infrared CN bands, 
and the 11.2\,$\mu$m band of SiC. They can be classified into several sub-types from 
the spectral features \citep{kee93}. \CR\ stars often show strong continuum in blue, 
slightly enhanced Ba, and strong isotopic C bands. \cn\ stars show strong and diffuse blue absorption, 
enhanced Ba lines, and weak isotopic C bands. \ch\ stars have strong CH absorption and weak Ca, 
Fe lines \citep{bar96}. Beside these, there are other sub-types such as \cj\ 
(strong $^{13}$C/$^{12}$C ratio) and dC (dwarf carbon) stars. Note that these classifications are not 
directly related to their origin.

Among these spectral sub-types, \cn\ and \ch\ stars are of more interest because the former are 
mostly in the AGB stage and the later are found mostly from halo populations \citep{gos05}. 
Increasing the samples for either types of carbon stars will be very helpful to improve our knowledge 
about how a star ends its life (through C-N stars) and the formation history of the stellar halo (through C-H stars). 
Based on the low-resolution spectra of the Sloan Digital Sky Survey \citep[SDSS;][]{yor00}, 39 and 251 
faint (R\,$>13$) high-latitude carbon stars (FHLCs) were identified successively by \citet{mar02} and \citet{dow04}, 
respectively. Recently, \citet{gre13} retrieved 1220 FHLCs from SDSS DR7 through cross 
correlation of stellar spectra with the SDSS carbon star templates. \citet{si14} found 202 
new carbon stars using the label propagation algorithm from SDSS DR8. 

From the Large Sky Area Multi-Object Fiber Spectroscopic Telescope (LAMOST) pilot survey 
\citep{cui12,den12,zha12}, \citet{si15} identified 158 new carbon stars using the manifold ranking 
algorithm. In a survey, the fraction of carbon stars that are dwarfs depends
sensitively on the magnitude limit of the survey. The relatively brighter limiting magnitude
($r\sim18$) for the LAMOST survey tends to find more giants and AGB stars, whereas the SDSS  
spectroscopy survey ($r\sim21$) reveals many more dwarfs than giants (nearly 70\%).

In this paper, we identify more carbon stars from LAMOST DR2, 
which contains almost 4 million stellar spectra, based on the spectral line indices. The paper is organized 
in the following manner. In section 2, we briefly introduce the basic data from LAMOST DR2 and give a detailed 
description of the method to search for carbon stars. In section 3, we show the results of the 
classification of the carbon stars. In section 4, we summarize the distribution of different sub-types 
of carbon stars in line indices and multi-band photometric space. Finally, 
a short conclusion is drawn in section 5.

\section{Data}
\subsection{The LAMOST Data}\label{sect:data}

The LAMOST, also called Guo Shou Jing telescope, is a 4-meter reflective Schmidt telescope with 4000 fibers 
on a 20-square degree focal plane \citep{cui12,zha12}. 
The unique design of LAMOST enables it to take 4000 spectra in a single exposure to a limiting magnitude 
as faint as $r=$\,19 at the resolution R\,=\,1800. The LAMOST survey will finally obtain more than 5 million low 
resolution stellar spectra after its 5-year survey \citep{den12}. Previous works have shown that LAMOST 
observations are biased more to the giant than the dwarf stars, since the limiting magnitude is relatively brighter 
\citep{liu14a,wan15}. By the end of 2014, the LAMOST team released the DR2 catalog, 
which contains 4\,136\,482 targets including 3\,784\,461 stars. In DR2, there are about 0.5 million 
late type giant stars \citep{liu15a}, implying that they should observe quite lots of carbon stars.

\subsection{Line Indices}\label{sect:lineind}

\begin{figure}
	\plotone{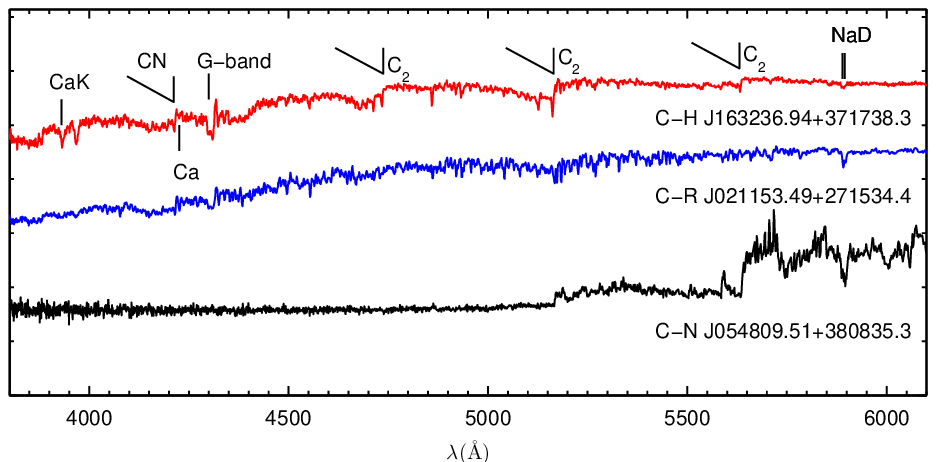}
	\plotone{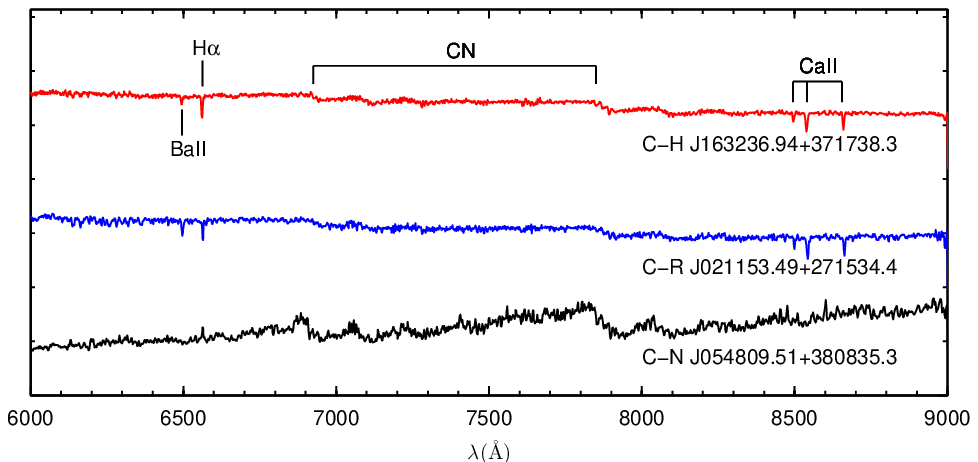}
	\caption{ LAMOST spectra for typical carbon stars.
 The features of the C$_2$ swan bands at 4737, 5165, 5635\,\AA\ and red CN molecule bands 
 at 7065 and 7820\,\AA\ are indicated. \label{fig:sample}}
\end{figure}

Typical spectra of carbon stars are characterized by strong C$_2$, CN, and CH absorption bands \citep{sec69}.
 Therefore, we measure the line indices of carbon related molecular lines to identify carbon stars. 
 We show several sample LAMOST spectra for typical carbon stars in Figure~\ref{fig:sample}. 
 The features of the C$_2$ swan bands at 4737, 5165, 5635\,\AA, and red CN molecule bands 
 at 7065 and 7820\,\AA\ are prominent in these spectra, while detailed features are different in 
 different sub-types. Compared to the traditional approach to classify the spectral types, 
 using line indices allows a semi-automatic classification. With multiple spectral line indices, 
 various spectral types may transitionally change from one to another, which naturally reflect 
 the variation of the astrophysical parameters, e.g. effective temperature, surface gravity, 
 and metallicity, from type to type \citep{liu15b}. 

In order to identify carbon stars, we adapt the Lick line indices defined by \citet{wor94} and 
\citet{wor97}. For the LAMOST DR2 spectra, we use the measured results from \citet{liu15b}. 
The Ba\,II (4554\,\AA), molecular bands \ctwo\ (around 5635\,\AA), CN (7065\,\AA), and CN (7820\,\AA) 
are also measured based on the wavelength definition listed in Table~\ref{tab:lineindices}.

The line index in terms of equivalent width (EW) is defined by the following equation \citep{wor94,liu15a}:
\begin{equation}
EW=\int(1-\frac{F_\lambda}{F_C})d\lambda,
\end{equation}
where $F_\lambda$ and $F_C$ are the fluxes of spectral line and continuum, respectively, which are functions of 
wavelength $\lambda$. The pseudo-continuum $F_C$ is estimated via linear interpolation of the fluxes located in 
the ``shoulder'' region on either side of each index bandpass. The line index under this definition is in \AA. 
For the spectra with signal-to-noise ratio larger than 20, the typical uncertainty of the equivalent widths of 
the atomic lines is smaller than 0.1\,\AA, while for the molecular band it becomes a few \AA, and may be systematically
affected due to difficulty in defining the continuum.

\section{Identification of carbon stars}

\begin{figure}
	\epsscale{1.10}
	\plotone{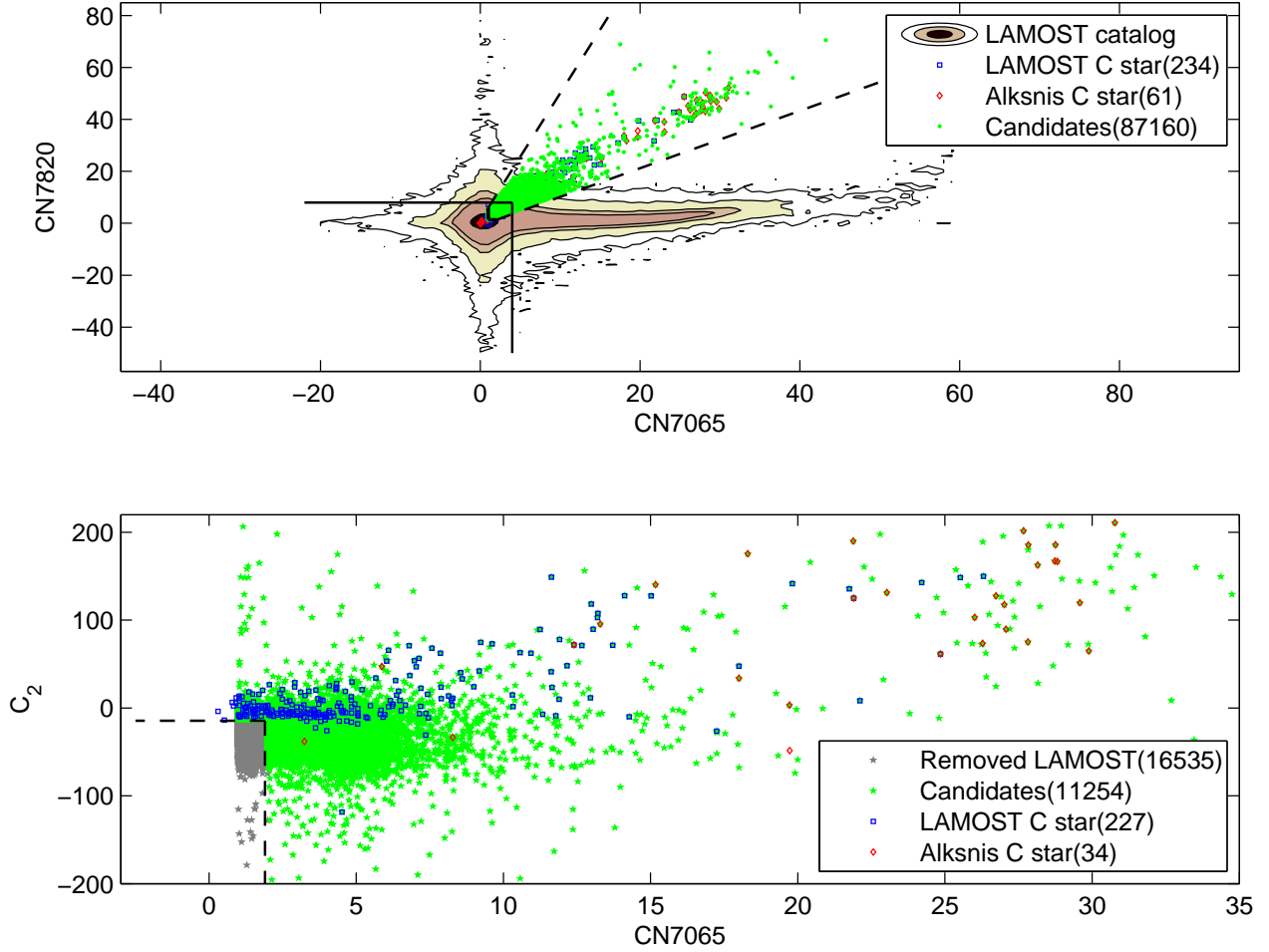}
	\caption{Distributions of the known carbon stars in CN (7820\,\AA) vs. CN (7065\,\AA)  (top panel) and in \ctwo\ (5635\,\AA) vs. CN (7065\,\AA) (bottom panel).
	The contours in the top panel represent the stellar spectra with S/N\,$>10$ in $i$ band from LAMOST DR2. The black dashed lines are the line cuts, which are 
	Equations (2)-(4) for the top panel and CN7065\,$=2$, CN7820\,$=-13$ for the bottom panel, and are used to select carbon star candidates (green filled asterisks).
	The blue unfilled squares represent the confirmed carbon stars with S/N\,$>20$ in $g$ and $i$ band from the 1093 
	stars with carbon flags in the LAMOST DR2 catalog, and the red unfilled diamonds represent the common carbon stars found by cross-identifying LAMOST DR2 (S/N\,$>20$ in both $g$ and $i$ bands) with the carbon star catalog provided by \citet{alk01}. The gray asterisks in the bottom panel represent the removed LAMOST stars. 
The two solid black lines are the borders that in their bottom-left region the additional cuts of S/N\,$>20$ in $g$ 
and $i$ bands for the stars are used in order to remove the spectra with low S/N in the central region. \label{fig1}}
\end{figure}

We first select the stellar spectra with signal-to-noise ratio (S/N) larger than $10$ in $i$ band from LAMOST DR2,
and obtain 2\,434\,289 stellar spectra in which only one epoch of the multiply observed objects is included.
Then we locate the distribution of the known carbon stars in \cnrt\ vs. \cnro\ plane (see the top panel of Figure~\ref{fig1}), 
so that we can determine the intrinsic location of the carbon stars. There are 1093 stars marked as carbon stars 
in the LAMOST DR2 catalog. These marks are added from a template matching technique \citep{luo15}. 
The automatic pipeline may mistakenly identify normal stars or those with lower S/N as carbon stars. 
Meanwhile, lots of the carbon spectra without a good match in their incomplete templates are missed. We inspect 
the stars with carbon marks by eye and finally confirm that 243 are real carbon stars with S/N$>20$ in both $g$ 
and $i$ bands (S/N($g$) and S/N$(i)$ hereafter). Moreover, we find 34 common carbon stars by cross-identifying LAMOST DR2 (S/N$>20$ in both $g$ 
and $i$ bands) with the carbon star catalog provided by \citet{alk01}.  

It is seen that, in the top panel of Figure~\ref{fig1}, the branch from the central region to the top-right corner is 
dominated by the carbon stars. 
We then adopt the following empirical criteria in the \cnrt\ vs. \cnro\ plane so that most of the stars located in the branch
toward top-right corner are included. With these cuts, we obtain 87\,160 carbon star candidates. 
\begin{equation}
48.5/44\times \mathrm{CN}7065-31/44<\mathrm{CN}7820<49/10\times \mathrm{CN}7065+11/10
\end{equation}
\begin{equation}
1<\mathrm{CN}7065<44
\end{equation}
\begin{equation}
1.5<\mathrm{CN}7820<71.
\end{equation} 

In the central region around the origin point of our selected branch in Figure~\ref{fig1}, different types 
of stars and low S/N spectra are mixed together. Hence, for the high contaminated region, 
it is very hard to identify the carbon stars in the crowded region only 
from the CN features. It seems that low S/N spectra are prominent in the crowded region. 
To remove them, we add additional cuts of S/N$>20$ in $g$ 
and $i$ bands for the stars located in the bottom-left region defined by the two solid black lines indicated in the figure. 
With this new cut, 59\,371 of 87\,160 stars are removed and 27\,789 carbon candidates are left.


We further map the candidate sample in the \ctwo\ vs. \cnro\ plane with the known carbon stars 
simultaneously cut with the same criteria in Equations (2)$-$(4), 
as shown in the bottom panel of Figure~\ref{fig1}. We remove the stars with weak \ctwo\ and \cnro\ absorption features, 
i.e. \ctwo\,$<-13$ and \cnro\,$<2$. Because the S/N is lower in the blue bands, in which most of the \ctwo\ bands
are located, it may be very difficult to identify \ctwo\ bands from the noisy spectra and lead to larger uncertainties.
Therefore, we decide to use CN features instead of bluer \ctwo\ features in this work. 
As a cost, some true carbon stars which have not CN features may also be removed after this cut. 
The new candidate sample (green symbols in the plot) includes 11\,254 stars. 
It is noted that the equivalent widths of \ctwo\ have lots of negative values, which are due to the variation of 
the pseudo-continua affected by the other molecular bands. This systematics is very difficult to be removed for 
late type spectrum since the molecular bands make it impossible to find the real continuum. Nevertheless, 
the systematic bias would not affect the identification of the carbon-rich features, because 
the relative values of \ctwo\ is sufficient in the identification.

\begin{figure}
	\plotone{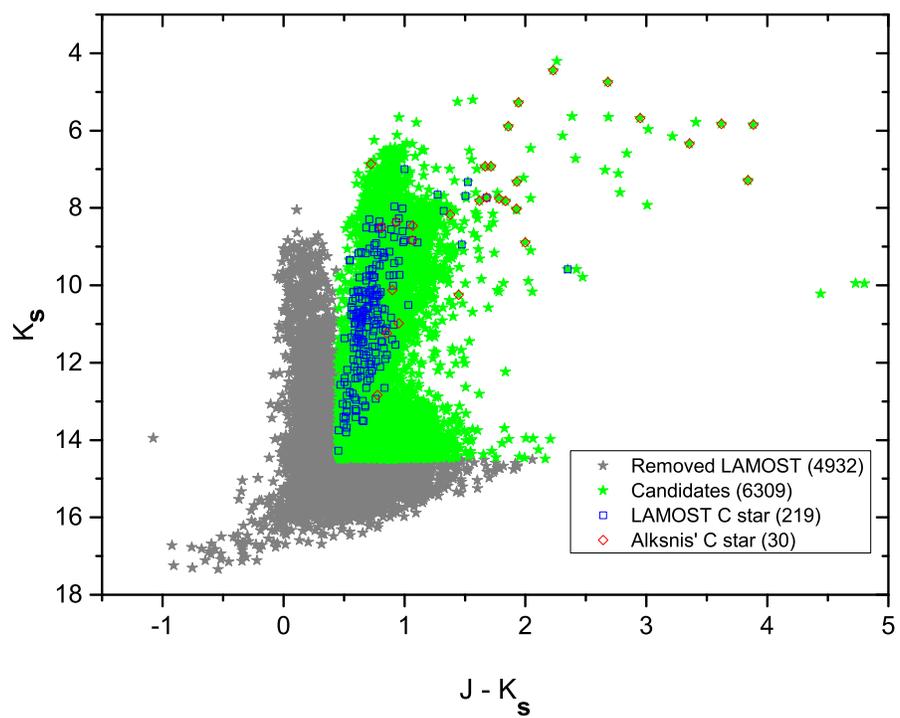}
	\caption{2MASS color-magnitude digram in $K_s$ vs. ($J-K_s$). The symbols are
		same as in Figure~\ref{fig1}.\label{fig:KJK}}
\end{figure}

These carbon star candidates are then mapped to the \K\ vs. \JK\ plane in Figure~\ref{fig:KJK}. Since 13 of these 11\,254 carbon 
star candidates have no $J$ and \K\ colors which were obtained, a total of 11\,241 candidates are shown in Figure~\ref{fig:KJK}. The known carbon stars 
are also overlapped in the same plot. The values of $J$, $H$, and \K\ are obtained by cross-identifying with the 
2MASS All-Sky point source catalog \citep{skr06}. 
As the limiting \K\ magnitude of 2MASS survey is around 14.5\,mag, we refine the sample with \K$< 14.5$\,mag 
so that the color index \JK\ is sufficiently accurate for the carbon star identification. In Figure~\ref{fig:KJK}, 
the stars brighter than \K$=14.5$\,mag can be divided into bluer and redder groups according to their \JK\ color. Most of 
the known carbon stars cover the redder group, which are dominated by the cool stars. With the cut of \JK$>0.45$\,mag, 
we can well remove most of the warm star contaminators located in the bluer group (marked as gray asterisks). From the cuts in 
the color-magnitude diagram, 6\,309 candidates are left. It is noted that some bluer carbon stars \citep[like the 
``G-types" in][]{gre13} may not be included. The ``G-type" stars are thought as a kind of dC star, 
which show weak CN and relatively blue colors and are probably the most massive dCs still cool enough to show \ctwo bands. 

Finally, we inspect the spectra for all 6\,309 stars by eye for confirmation. The CN, \ctwo, CH, and TiO$_2$ features 
are used to judge whether the candidate spectra are real carbon stars. This procedure has been done at least twice 
by the same people and has been independently confirmed by other people to avoid severe bias of the judgment. 
The whole selection process for carbon stars is listed in Table~\ref{tab:steps}, and
we finally identify 894 carbon stars, which are listed in Table~\ref{tab:carbonstars}. 

\begin{figure}
	\plottwo{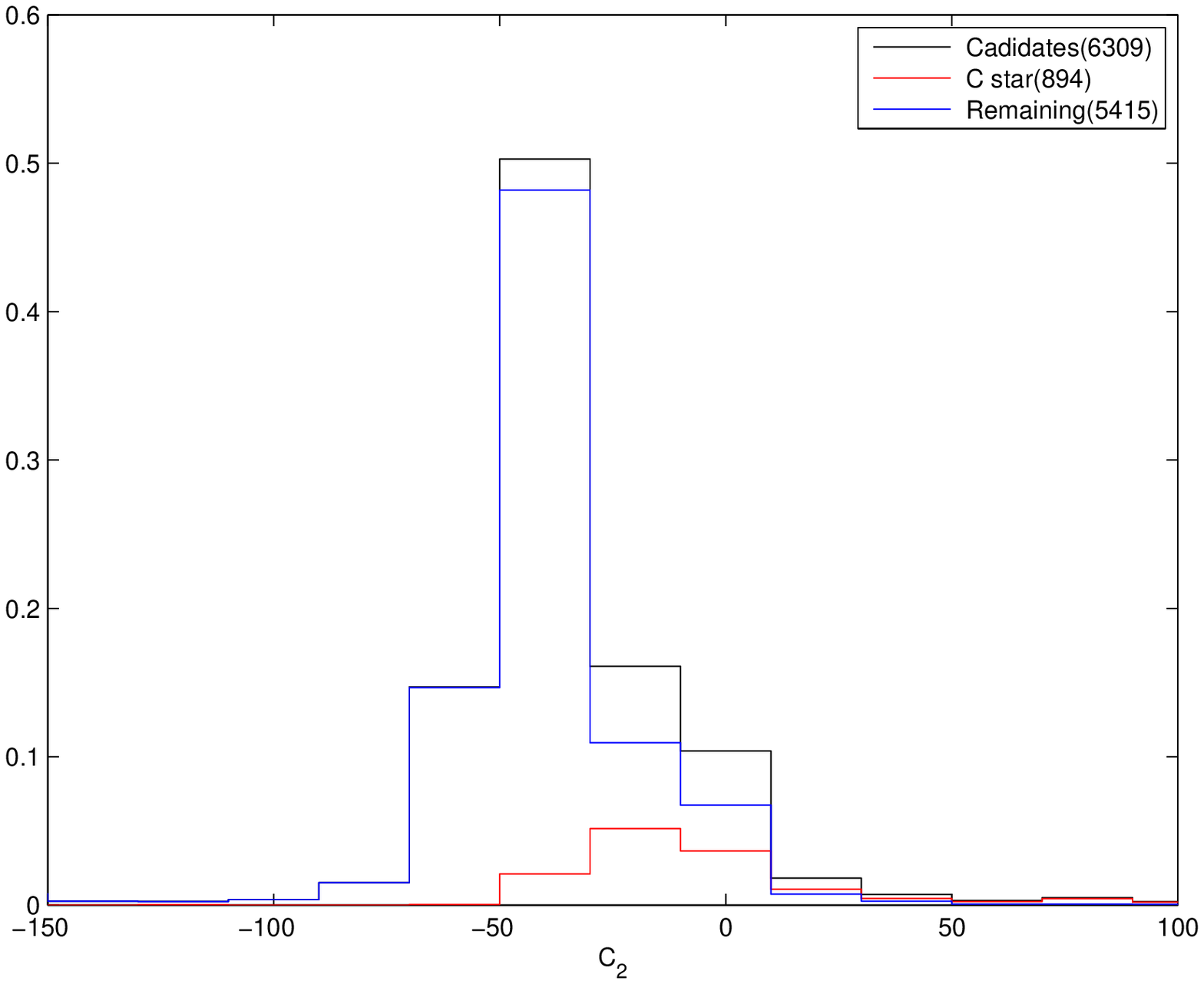}{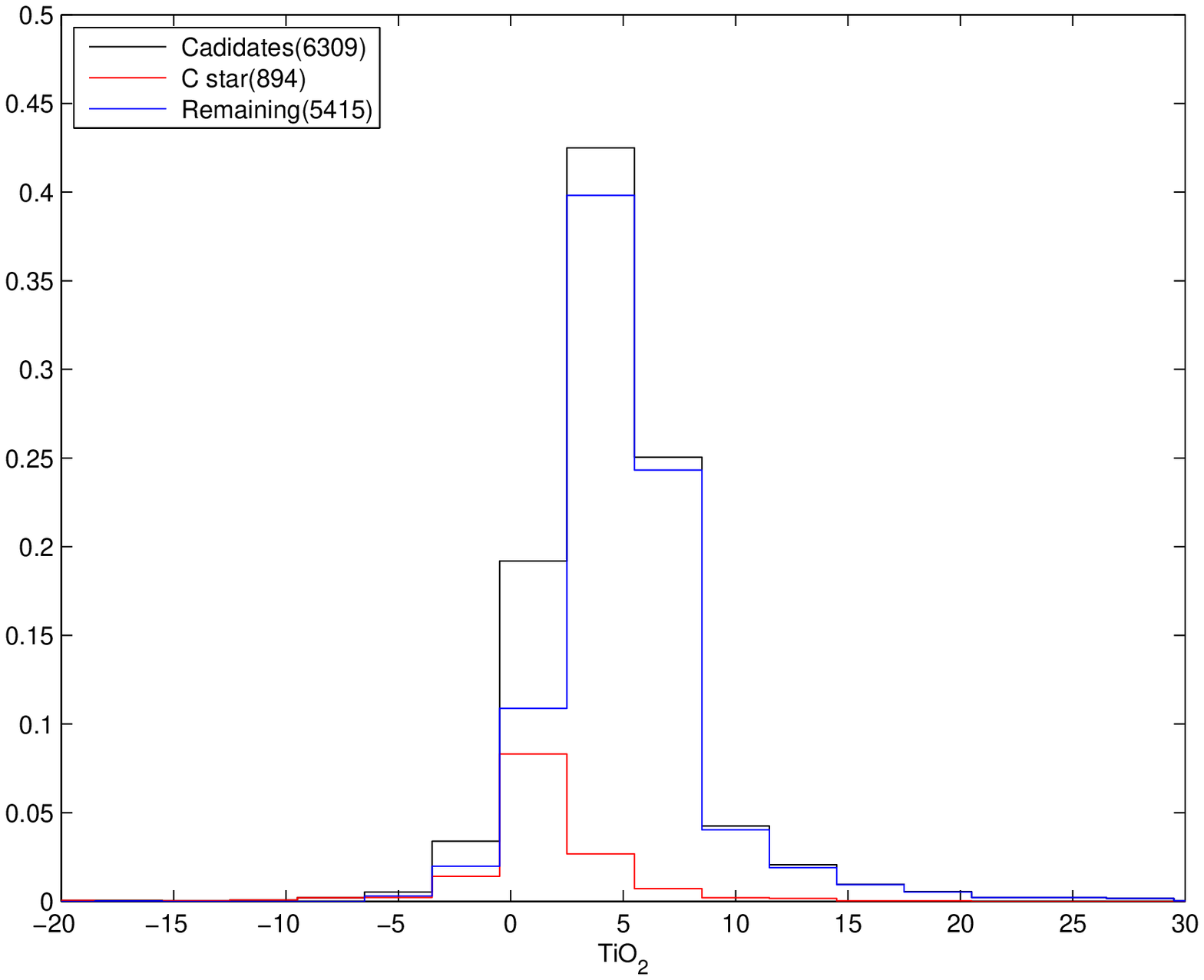}
       \caption{Normalized histograms of C$_2$ (5635\,\AA, a) and TiO$_2$ (6191\,\AA, b) for the identified 894 carbon stars (red), the remaining 5\,415 candidates (blue), 
       and the total 6\,309 candidates (black).\label{fig:tioc2}}
\end{figure}

Figure~\ref{fig:tioc2} shows the normalized histograms of EW for C$_2$ (5635\,\AA, left panel) and TiO$_2$ (6191\,\AA, right panel) 
for the 894 carbon stars, the remaining 5\,415 candidates, and the total 6\,309 candidates. 
It is noted that many of (but not exclusively) the visually-selected carbon stars are found in the region of strong \ctwo\ 
($>-30$) and weak TiO$_2$ ($<2.5$). However, some unselected candidates are also located in 
the weak-TiO$_2$, strong-\ctwo\ region. They are double checked and confirmed that they are contaminators due to 
the noise in the spectra. Some of the weak-TiO$_2$ stars also show strong \ctwo, most of them are not carbon 
stars but their \ctwo\ measures are affected by the local spikes in the spectra.

\begin{figure}
	\plotone{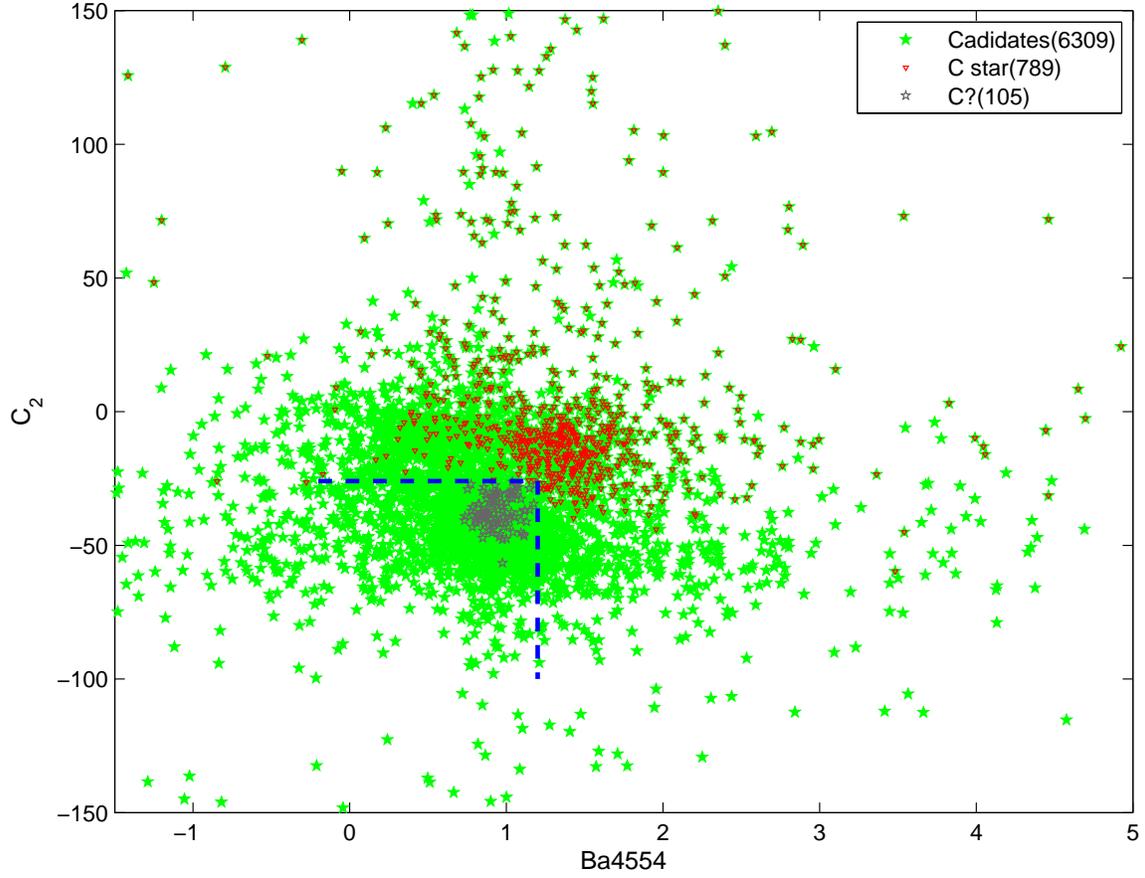}
	\caption{Distributions of identified carbon stars in this work (red unfilled triangle) in the C$_2$ (5635\,\AA) vs. Ba\,II (4554\,\AA) plane. 
	The gray unfilled asterisks represent the suspect carbon stars, and the green filled asterisks represent the candidates from LAMOST DR2. 
	The blue dashed lines are the line cuts (C$_{2}=-16$, Ba\,II(4554\,\AA)\,$=1.2$) by which we define the suspect carbon stars. \label{fig:bac2}}
\end{figure}

\begin{figure}
	\plotone{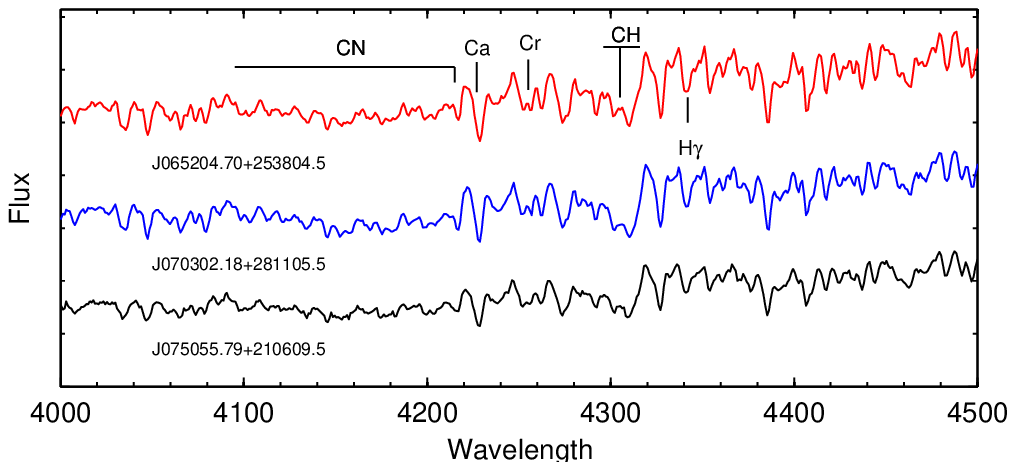}
	\plotone{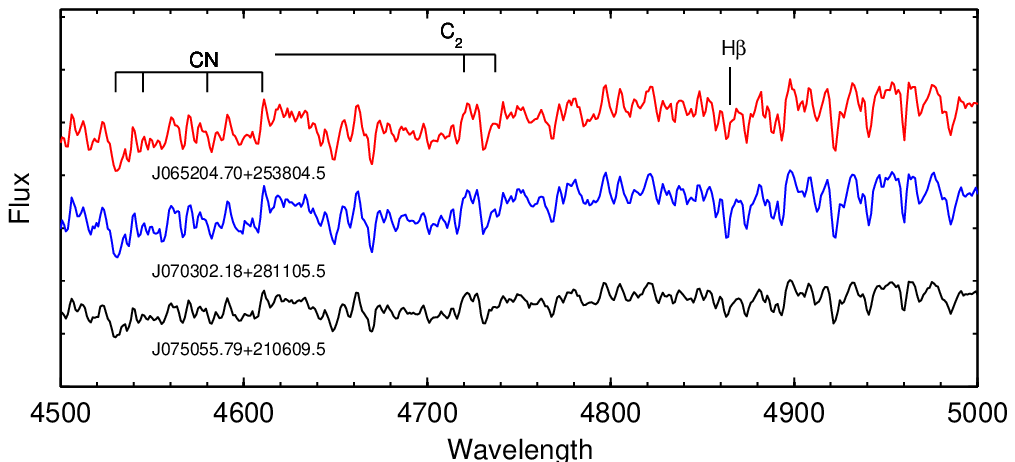}
	\plotone{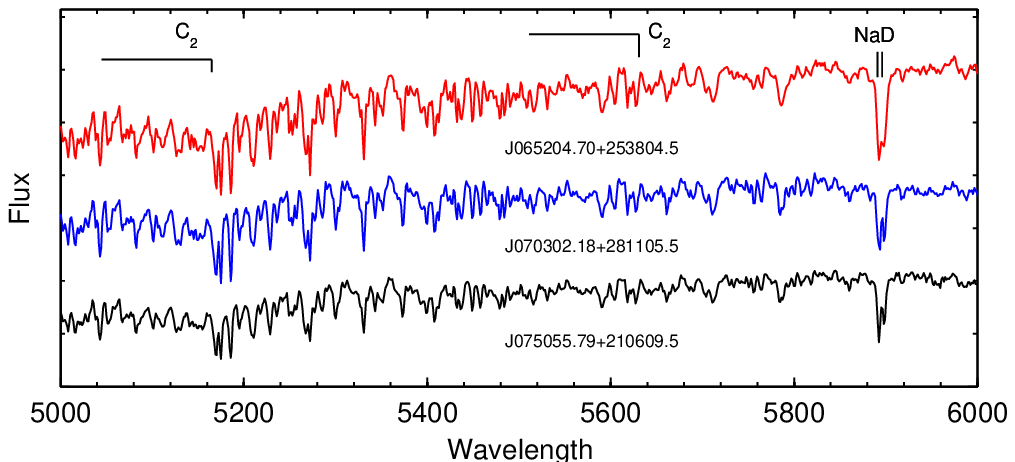}
	\caption{Three sample spectra for the suspected carbon stars located at the bottom-left corner of Figure~\ref{fig:bac2}.\label{fig:suspectC_spectra}}
\end{figure}

We further map the carbon stars \ctwo\ vs. Ba4554)  in Figure~\ref{fig:bac2} and surprisingly find some carbon stars are located in 
a clump mostly occupied by the ejected O-rich stars. We select the clump with two straight-line cuts 
at \ctwo$=-16$ and Ba\,II (4554\,\AA)=1.2 (blue dashed lines shown in the plot) and obtain 105 samples located in the clump. 
We double check these samples and find that they do show \ctwo\ in their blue spectra and CN band in the
red, which are very likely features of early type carbon stars. Figure~\ref{fig:suspectC_spectra} shows a 
few sample spectra selected from this group of stars. Because in this work we do not directly measure C/O, which should be 
determined from the comparison with a set of proper model spectra, it is not clear whether the C/O ratio is 
larger than 1 for these stars. Therefore, we mark them in the Table~\ref{tab:carbonstars} with a question flag and denote them 
as the \emph{suspected carbon stars's in the rest of the paper}.
Alternatively, 
these stars may be the so called strong-CN stars  \citep{gra09}. 
More investigation for these stars should be done in future work.

\section{Sub-classification}

\subsection{Criteria for spectral sub-type}

According to the revised MK carbon-star classification system of \citet{kee93}, there are five types of carbon stars, 
i.e., \CR\ stars, \cn\ stars (roughly corresponding to the Harvard R and N stars), \cj\ stars, \ch\ (or CH) stars, and 
C-Hd (hydrogen deficient) carbon stars. Here, we mainly focus on the types of \CR, \cn, and \ch.

\cn\ stars can be easily distinguished from other types because of the strong absorption in the blue part of the
spectra~\citep[generally little or no flux bluer than 4400\,\AA\ according to][]{kee93}. On the other hand, \ch\ stars 
are difficult to be distinguished from \CR\ stars. In general, \ch\ stars are more metal poor and most of them are 
recognized as high-velocity halo objects. However, there are still some indicators which can be used to distinguish \ch\ stars 
from \CR\ stars based on the low-resolution spectra, such as 1) \ch\ stars show quite a strong G band; 
2) they have an exceptionally strong P-branch head near 4342\,\AA; 3) they show weaker Ca\,I at 4227\,\AA; 
4) they show enhanced lines of s-process elements (e.g., Ba\,II lines at 4554, 5853 and 6496\,\AA) 
and weaker Fe-group elements \citep[][]{kee93,bar96,si15}.


According to the low-resolution LAMOST spectra, we select a set of the most prominent criteria, which are summarized in
Table~\ref{tab3},  to classify the sub-types of the carbon stars. Since we do not use any quantified approach 
in the procedure, misclassification may occur for a few carbon stars with insignificant features, 
especially for those in between \CR\ and \ch.
Applying the criteria listed in Table~\ref{tab3}, we find 108 \cn, 259 \CR, 339 \ch, and 83 unknown stars. 
The sub-type for each carbon star is listed in Table~\ref{tab:carbonstars}.

After manually classifing the carbon stars, we reveal more features of the sub-types in their spectral and photometric data. 
On one hand, these features will be very helpful to better understand the physics of these stars; on the other hand, 
they can provide some hints for improving the approaches to the sub-classification. In next two sub-sections, 
we discuss the features for the \cn\ stars and \CR/\ch\ stars, respectively.

\subsection{\cn\ stars}
\begin{figure}
	\plotone{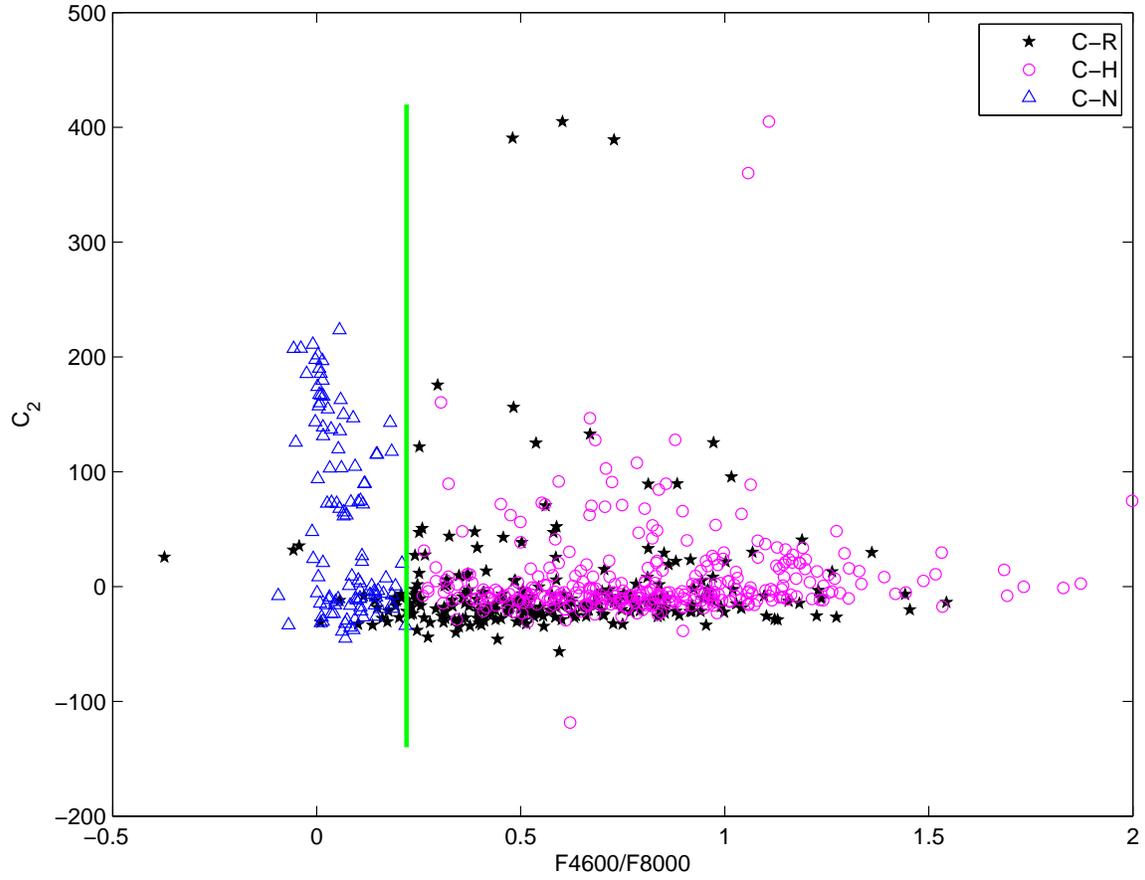}
	\caption{Distributions of the identified C-N (blue unfilled triangles), C-H (meganta unfilled circles) and C-R (black asterisks) in C$_2$ (5635\,\AA) vs. F4600/F8000 plane. The green line is F4600/F8000\,$=0.22$, at which we can separate all 
	of the C-N stars from the C-H and C-R stars. \label{fig:c2_blueflux}}
\end{figure}

Among the three sub-types discussed in this work, \cn\ stars are the easiest one to be classified from the spectra, 
since they show very weak fluxes in blue wavelengths. Hence we develop a color index, denoted as F4600/F8000, 
by dividing the median flux between 4300 and 4600\,\AA\ by the value between 8000 and 8700\,\AA. 
We do not use fluxes bluer than 4300\,\AA\ because they are more affected by noise. 
Figure~\ref{fig:c2_blueflux} shows that most of the \cn\ stars are located in the left side with very low flux in 
blue wavelength. It also displays that the equivalent width of \ctwo\ is slightly anti-correlated with F4600/F8000, 
implying that the \cn\ stars with stronger swan bands have lower flux in the blue end. 
This trend can be naturally explained by the effective temperature. The lower effective temperature, 
the stronger the molecular bands and the lower luminosity in blue band.

\begin{figure}
	\plotone{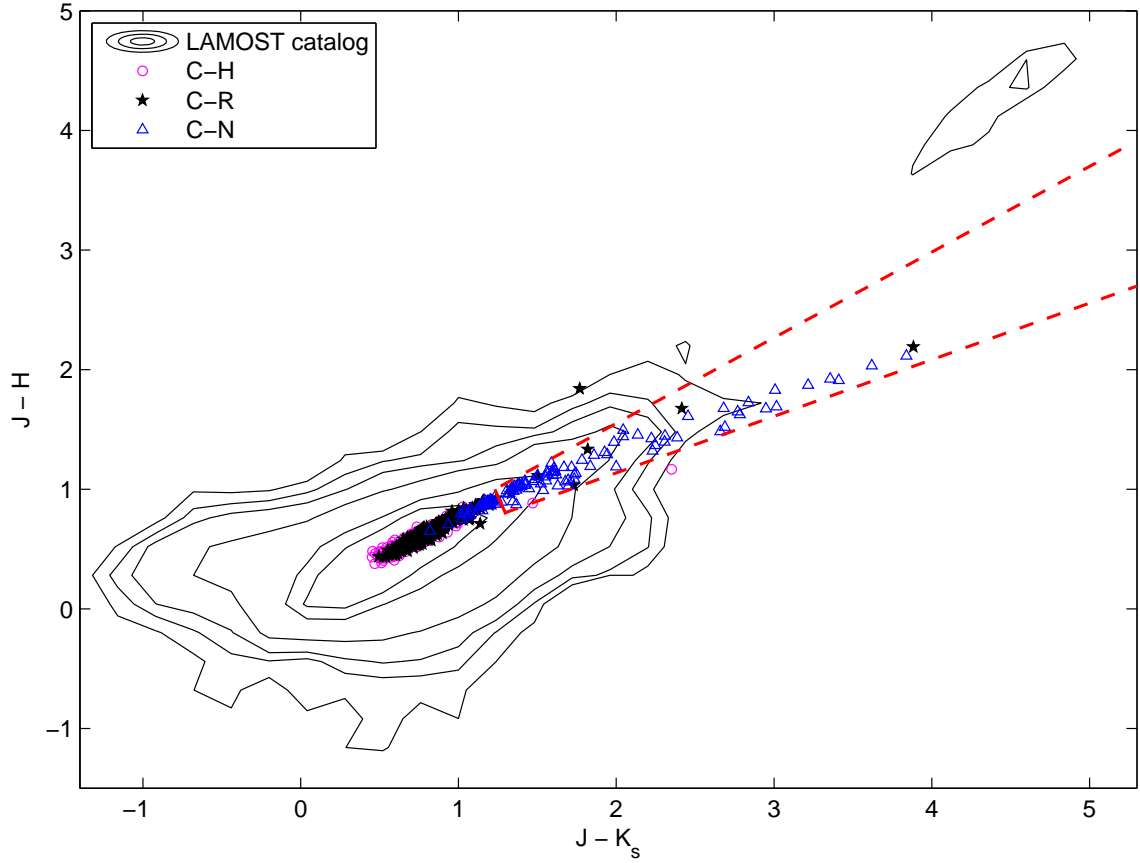}
	\caption{Stellar locus in the \JH\ vs. \JK\ diagram. The red dashed lines are the line cuts, which are 
	\JK\,$>49\times(J-H)/35.1-5.8/35.1$, \JK\,$>5/3.25-(J-H)/3.25$, \JH\,$>0.5\times$(\JK)$+0.15$, and are used to select C-N stars. The contours indicate 
	the location of the normal stars from the LSMOST DR2 catalog. The rest symbols are same as in Figure~\ref{fig:c2_blueflux}.\label{fig:ccdiagram}}
\end{figure}

Lots of previous work has demonstrated that the infrared photometry is very helpful to identify the \cn\ stars. 
Figure~\ref{fig:ccdiagram} shows that the \cn\ stars are concentrated in the red end of the stellar locus 
in \JH\ vs. \JK\ diagram, specifically redder than the boundary defined by \JK\,$>49\times(J-H)/35.1-5.8/35.1$, \JK\,$>5/3.25-(J-H)/3.25$, and \JH\,$>0.5\times$(\JK)$+0.15$.

Figures~\ref{fig:c2_blueflux} and~\ref{fig:ccdiagram} provide two different means to identify \cn\ stars. 
Based on our sample, for the low resolution spectroscopic data, if we classify the \cn\ as those bluer than $F4600/F8000<0.22$, 
the completeness of this simplified criterion can reach 100\% with 16\% contamination. For the photometric data, 
if we apply the criteria of \JK\,$>49\times(J-H)/35.1-5.8/35.1$, \JK\,$>5/3.25-(J-H)/3.25$, and \JH\,$>0.5\times$(\JK)$+0.15$, we can identify 73\% from the full sample of the \cn\ stars with a contamination rate of 8\%.

\subsection{\CR\ and \ch\ stars}
\begin{figure*}
	\begin{minipage}{18cm}
		\plottwo{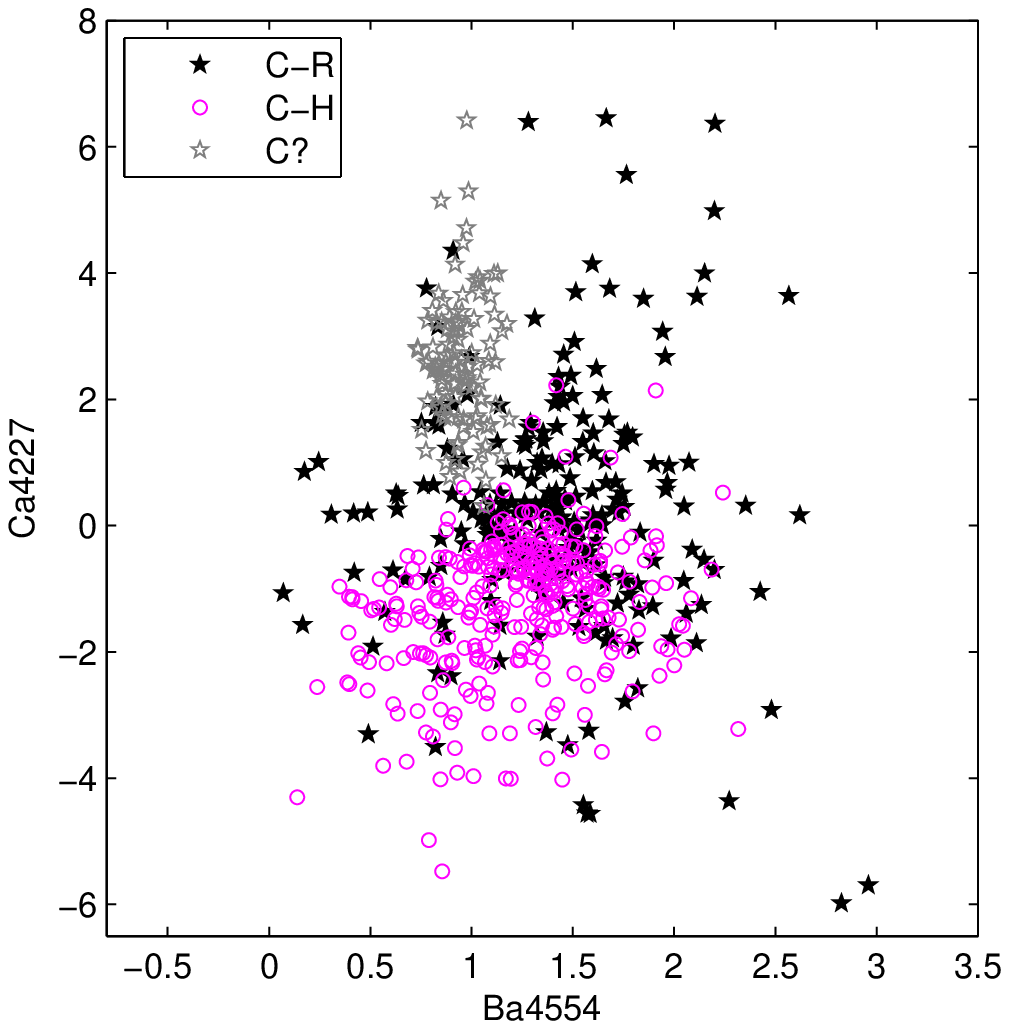}{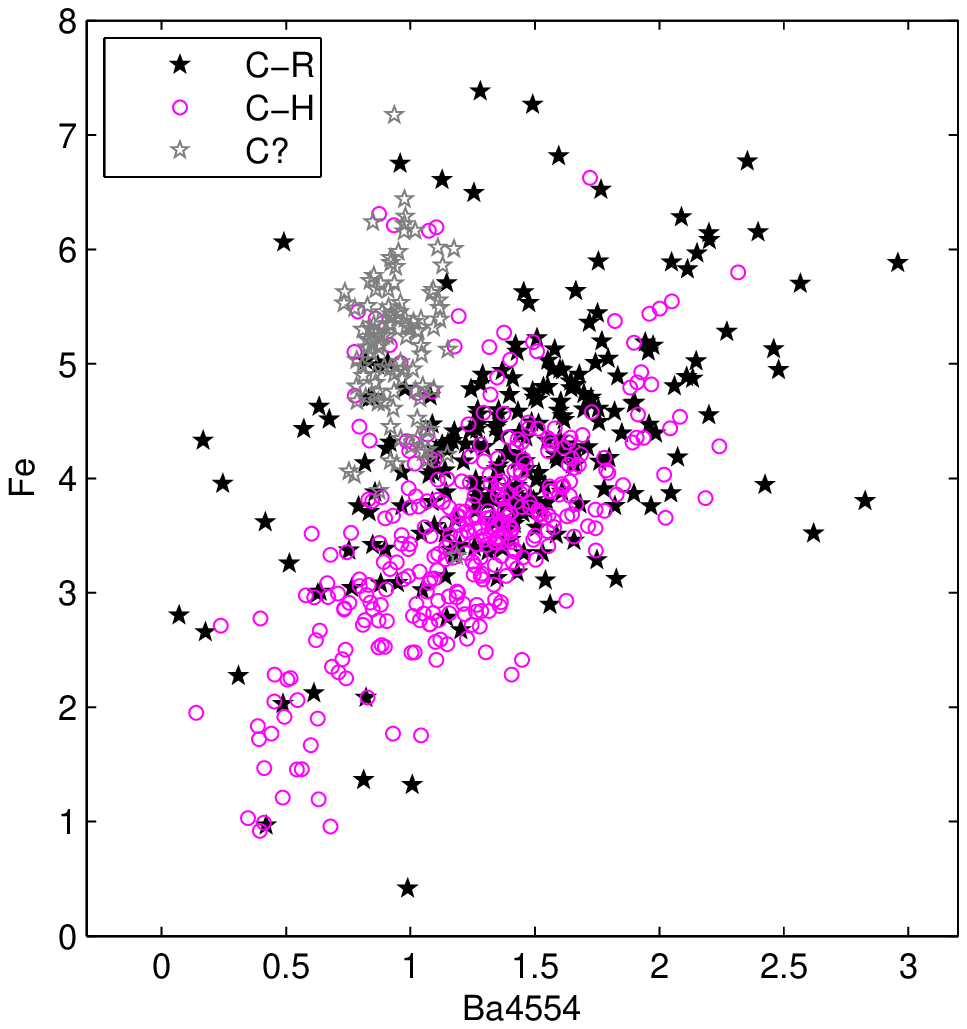}
	\end{minipage}
	\caption{Distributions of the \ch\ (magenta unfilled circles), \CR\ (black filled asterisks), and the suspect carbon (gray unfilled asterisks) stars. 
	The left panel shows the Ca\,I (4227\,\AA) vs. Ba\,II (4554\,\AA) plane and the right panel shows the Fe vs. Ba\,II (4554\,\AA) plane.\label{fig:metal}}
\end{figure*}

\begin{figure}
	\plotone{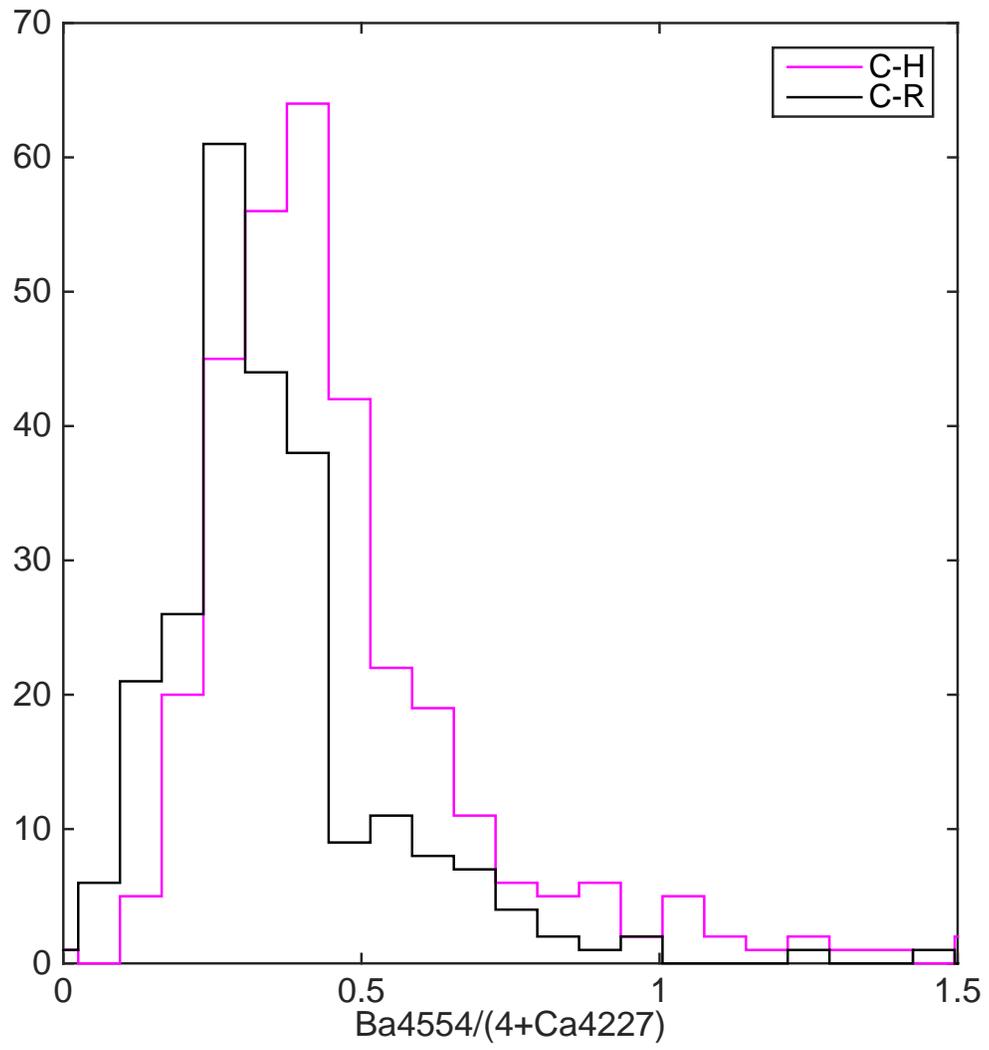}
	\caption{Histogram of the ratio of Ba\,II\,(4554\,\AA) to Ca\,(4227\,\AA) for \ch\ (magenta) and \CR\ (black) stars.\label{fig:hist_BaCa}}
\end{figure}

We show the distributions for the \CR\ and \ch\ stars in the Ba\,II (4554\,\AA)-
Fe\footnote{The averaged EW for nine Fe Lick line indices located between 4000 and 6000\,\AA.}-Ca\,I (4227$\rm\AA$) 
space in Figure~\ref{fig:metal}. It is seen that, although on average the \CR\ are more metal-rich than the \ch\ stars, 
no clear boundary is found between the two classes.
The suspected carbon stars (gray unfilled asterisks) 
are also drawn in the figure. The left panel shows that most of the \ch\ stars have similar Ba but lower Ca 
compared with the \CR\ stars. The right panel shows that the equivalent width of Ba for both \CR\ and \ch\ stars 
are well correlated with Fe, while the suspected carbon stars are isolatedly located above the \CR\ and \ch\ stars with 
larger Fe. It is not easy to directly estimate whether the \ch\ stars are more Ba-rich than the \CR\ stars simply 
from the line indices, because the line indices are not only affected by the abundance but also 
correlated with effective temperature, surface gravity log\,$g$, and overall metallicity. Therefore, we use the lines indices ratio Ba4554/(4+Ca4227) as the indicator of the [Ba/Fe], 
since the Ca line can be approximately treated as a proxy of the metallicity without being sensitively affected by 
the effective temperature. However, the Fe lines are more sensitive to the effective temperature and hence 
cannot be directly used as a proxy of the metallicity. Figure~\ref{fig:hist_BaCa} shows that, although 
they are quite similar, the Ba abundance for the \ch\ stars is statistically larger than that for the \CR\ stars.

\begin{figure}
	\plotone{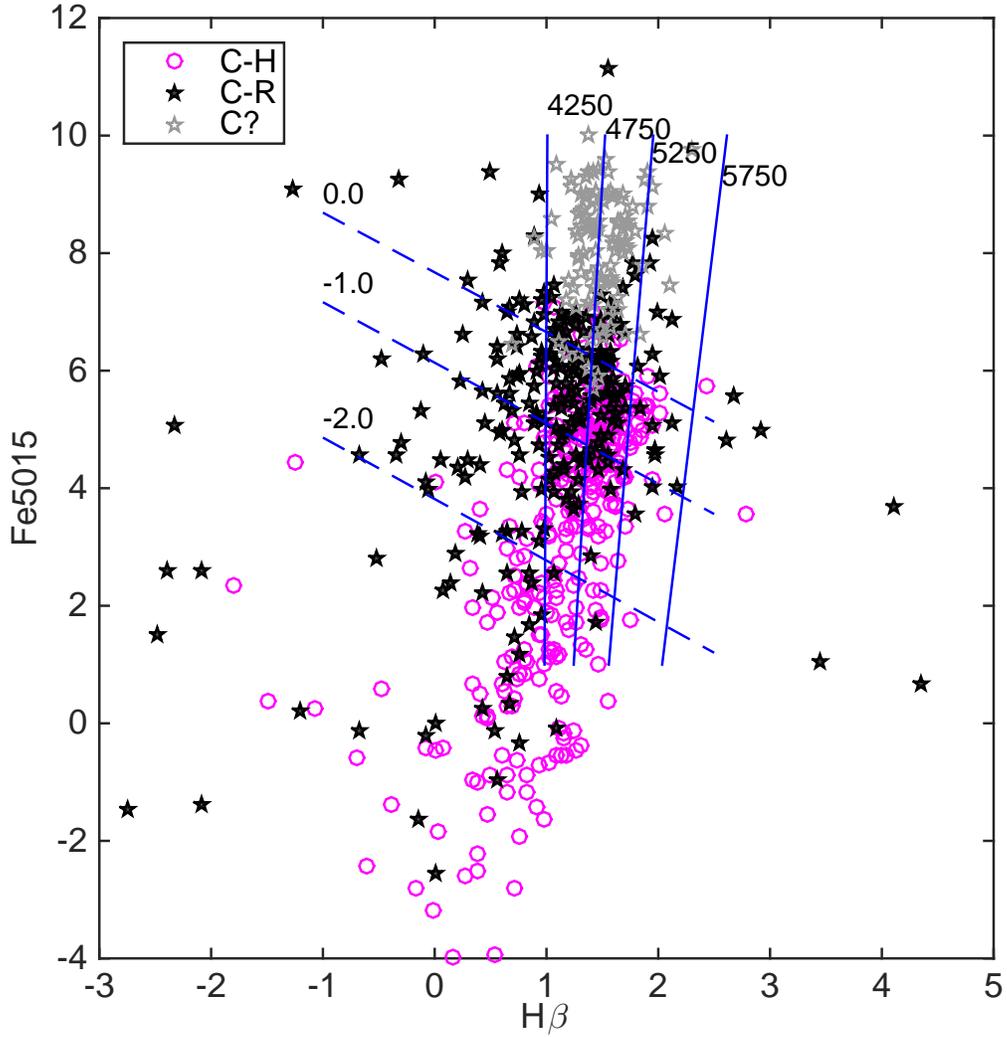}
	\caption{Distribution in Fe (5015\,\AA) vs. H$\beta$ for \ch\ and \CR\ stars. The blue solid lines indicate the effective temperature grid (4250, 4750, 5250, and 5750\,K), empirically derived from the normal K giant stars from LAMOST DR2 and the dashed blue lines indicate the metallicity grid (-2, -1, 0\,\,dex) derived in a 
    similar way to the effective temperature grid. 
	The symbols are same as in Figure~\ref{fig:metal}. \label{fig:hbeta_fe5015}}
\end{figure}

In Figure~\ref{fig:hbeta_fe5015}, the \CR\ and \ch\ stars are compared with the normal K giant stars in effective 
temperature and metallicity via two line indices: $H_{\beta}$, which is sensitive to the effective temperature, 
and Fe (5015), which is sensitive to the metallicity \citep{liu14b}. We find that the \CR\ are more metal-rich and slightly 
cooler than \ch\ stars. Interestingly, the suspected carbon stars (the gray unfilled asterisks) are mostly supersolar 
metal-rich stars.

The \ch\ stars are believed to be in binary systems, in which the companions have evolved to white dwarfs and mass 
transferring may be responsible for the enrichment of the carbon at the surface of the \ch\ stars. 
The origin of the \CR\ stars, especially the early type \CR\ stars, is still not quite clear. 
It seems that they are not post-AGB stars due to the lower luminosity and some observational evidence shows that 
they are also not mass transfer binaries \citep{mcc97}. Although \citet{mcc97} suggested that the \CR\ stars may have been coalesced, 
\citet{wal98} thought no additional rotation signature is found from the width of the spectral lines for 
the sample of McClure. Theoretical works suggested that a merger of a helium white dwarf to a red giant branch 
star can shift the helium-flash to the outskirt of the core allowing more carbon elements to mix in 
the convective envelope \citep{izz07,dom10}, while another scenario of red giant mergers may not create 
a peculiar helium-flash and additional carbon at surface \citep{pie10}. The larger Ba abundance of 
\ch\ stars is likely due to
mass transfer from an AGB companion. The smaller Ba for \CR\ stars is consistent with the 
hypothesis that they should not be polluted by an AGB companion.  

\subsection{Distributions}

\begin{figure}
	\epsscale{1}
	\plotone{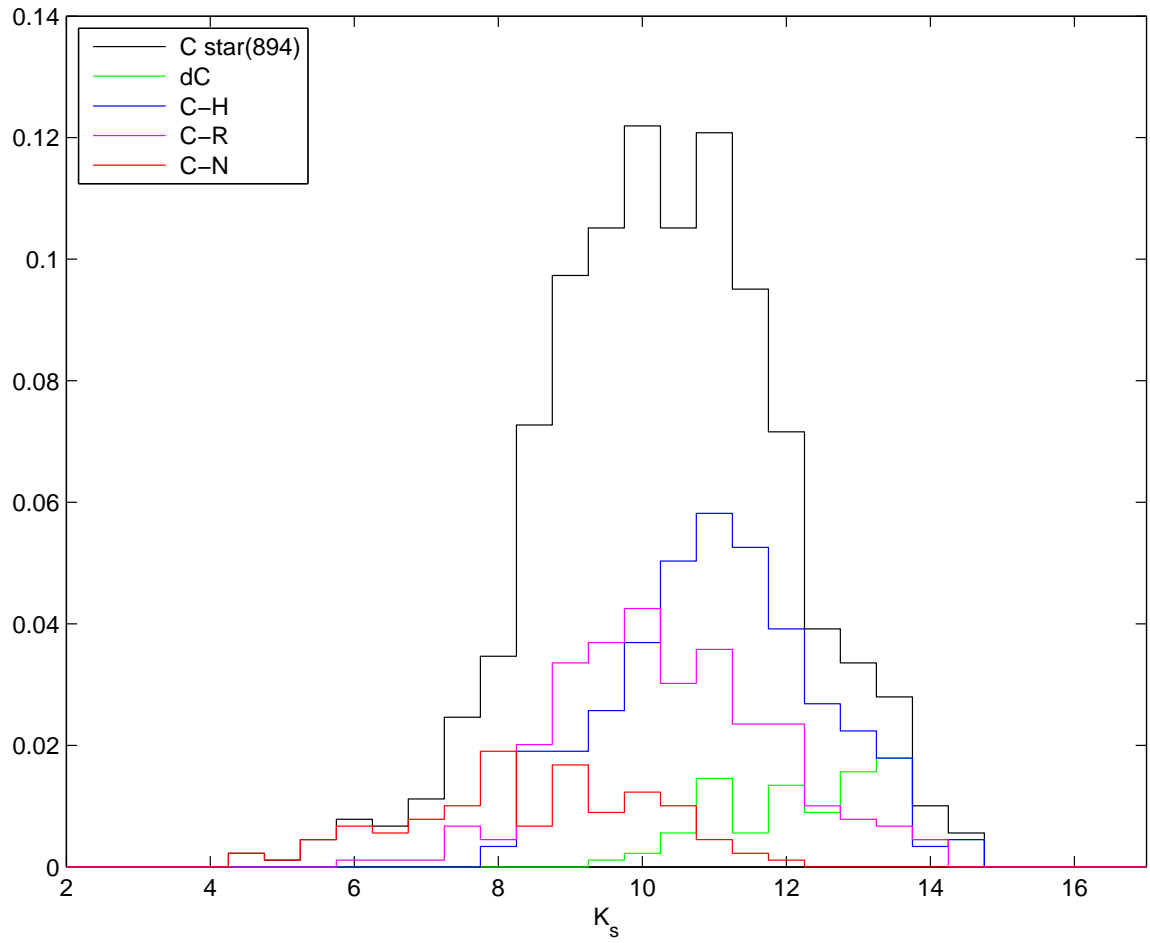}
	\caption{Histograms of magnitude $K_s$  for dC (green), C-N (red), C-H (blue), C-R (magenta), and the total number of carbon (black) stars.\label{fig:ks}}
\end{figure}

Figure~\ref{fig:ks} shows histograms of magnitude $K_s$ for dC, \ch, \CR\ and \cn\ stars. It can be seen that
the \cn\ stars are brightest, then \CR, \ch, and dCs the last, which is expected. Since \cn\ stars are N-type AGB stars
with high luminosity, they show very bright $K_s$ magnitude. dCs are dwarf stars with the lowest luminosity, thus they
concentrate in the faintest end. Most of \CR\ and \ch\ stars are giants, thus have 
moderate distributions of magnitude. The identification for dCs is discussed in section 5.

\begin{figure}
	\epsscale{1}
	\plotone{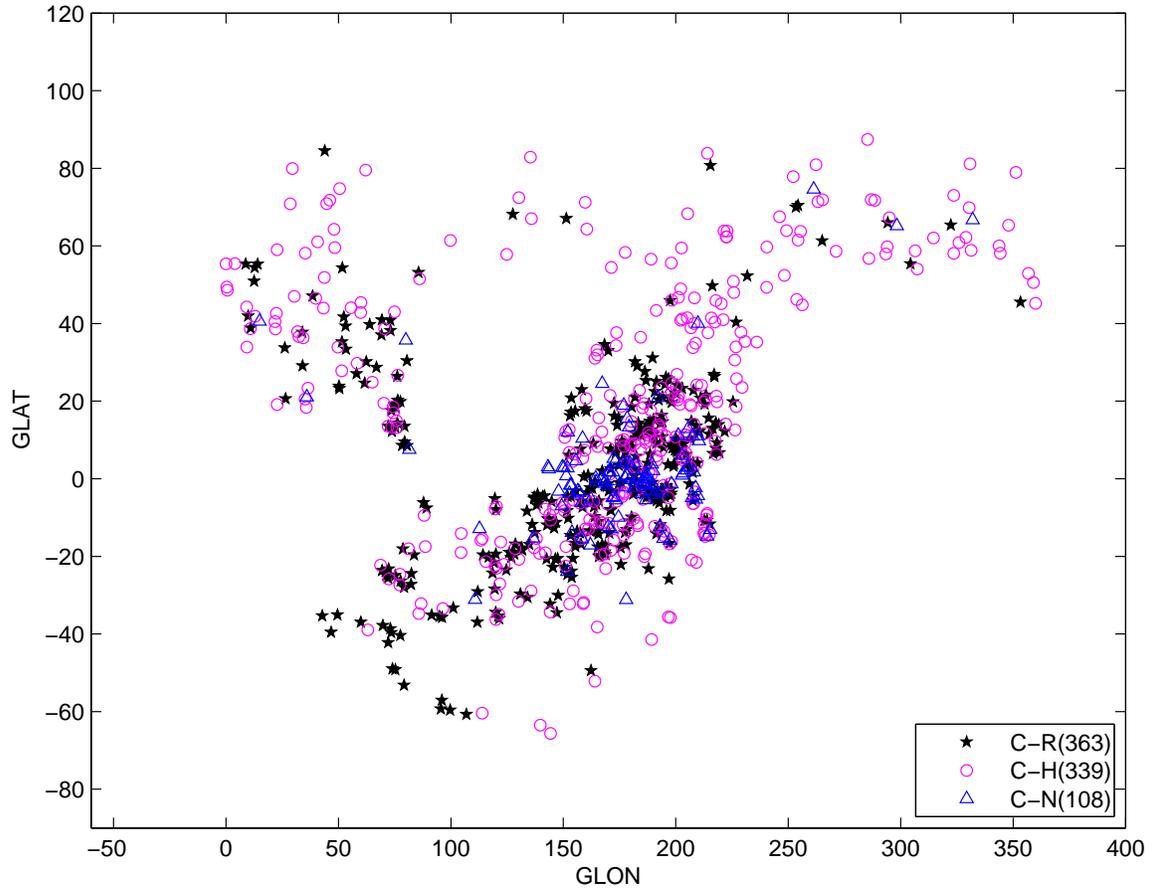}
	\caption{Spatial distribution of C-N, C-H and C-R stars. The symbols are same in Figure~\ref{fig:c2_blueflux}.\label{fig:spatial}}
\end{figure}

Finally, Figure~\ref{fig:spatial} shows the spatial distribution for the identified carbon stars in Galactic coordinates. 
As expected, the \cn\ stars are quite concentrated in the disk mid-plane, while the \ch\ stars are mostly distributed 
in the high Galactic latitude region. The \CR\ stars are in the middle, most of them are concentrated at low Galactic latitudes 
but a few are located at high latitudes.

\section{Discussion and conclusion}
\begin{figure*}
	\epsscale{1}
	\plotone{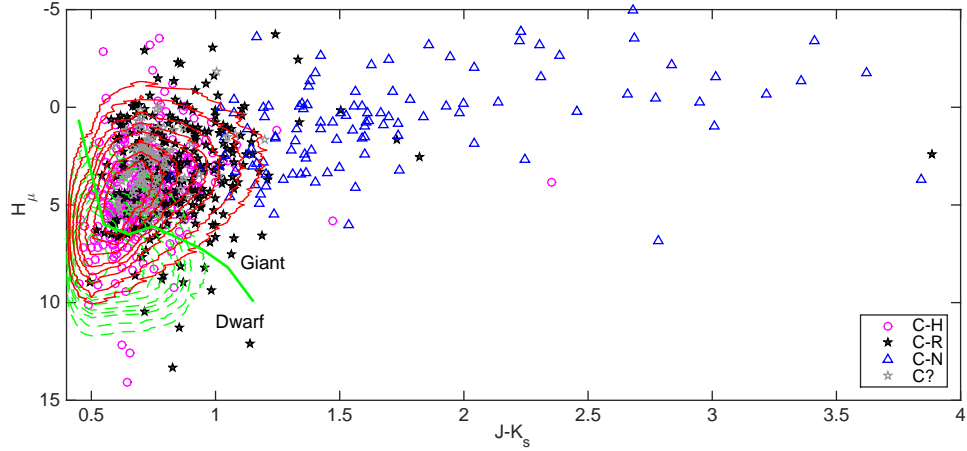}
	\caption{Reduced proper motion (H$_\mu$) vs. 
\JK\ for the carbon stars as well as the distributions for the normal K giant stars (defined as log\,$g<3.5$ and 
displayed as the red contours) and K dwarf stars (defined as log\,$g>4$ and displayed as the green dashed contours) 
selected from the LAMOST DR2 catalog. The solid thick green line indicates the middle points between the two probability density functions, 
which can be used as the separation line for the giant/dwarf stars. \label{fig:redpm}}
\end{figure*}

Previous works showed that many of the carbon stars are actually the carbon dwarf stars. However, 
it seems very difficult to distinguish dwarf stars from the spectral features. Hence, we identify 
the possible dwarf stars from the reduced proper motion. The reduced proper motion is defined as
\begin{equation}\label{eq:redPM}
H_\mu=K_s+5\mathrm{log}\sqrt{\mu_\alpha^2+\mu_\delta^2}-10,
\end{equation}
where $H_\mu$ is the reduced proper motion, \K\ is the $K$-band photometry from the 2MASS catalog, $\mu_\alpha$ and $\mu_\delta$ 
are the proper motion in mili-arcsecond per year from the PPMXL catalog \citep{roe10}. Because most of the stars 
in the Milky Way have relatively similar motion speed, then the transversal angular velocity, i.e. the proper motion, 
is generally larger (smaller) when the star is nearer (farther) to the Sun.  Consequently, the proper motion can 
be very roughly considered as the proxy of the parallax and thus $H_\mu$ can be roughly treated as the absolute 
magnitude of a star according to Equation (\ref{eq:redPM}). Figure~\ref{fig:redpm} shows $H_\mu$ vs. 
\JK\ for the carbon stars as well as the distributions for the normal K giant stars (defined as log\,$g$\,$<3.5$ and 
displayed as the red contours) and K dwarf stars (defined as log\,$g$\,$>4$ and displayed as the green dashed contours) 
selected from the LAMOST DR2 catalog. The two distributions for the normal K giant and dwarf stars can be used as 
the probability density functions for the two luminosity types in the plane and the middle points between the two probability density functions 
are indicated by the solid thick green line, which can be used as the separation line for the giant/dwarf stars. 
The stars located above the line are very likely giant stars and the stars located below are likely dwarf stars. 
Applying this empirical giant/dwarf separation line to the 894 carbon stars, we find 84 possible carbon dwarf stars, 
53 of them are \ch\ stars and 22 of them are \CR\ stars. None of the possible dwarf stars are classified as the \cn\ or suspected carbon stars.

We investigate the 243 confirmed carbon stars with \emph{carbon} flag in the LAMOST catalog. Where 219 of them are left after the cuts 
in \cnrt, \cnro, \ctwo, \K, and \JK. When we inspect the spectra for the 6\,309 candidates we do not check the \emph{carbon} flag. 
But finally, all of the 219 confirmed LAMOST carbon stars are again identified as the carbon stars and show up in the final catalog. This confirm that the inspection of the spectra is quite
stable and reliable.

In this work, we combine the manual inspection of the spectra with the cuts in the line index space and color-magnitude diagram, and 
successfully identify 894 carbon stars from the LAMOST DR2 catalog. These 105 of the suspected samples
have weak \ctwo\ band with weak TiO$_2$ band, which is not a typical feature of carbon stars.  
However, to keep the catalog complete, we leave them in the final catalog with a mark of type \emph{C?}. The  largest fraction of the stars, 
i.e. 339 of 894, are classified as the \ch\ stars, which are Ba enhanced and more metal-poor. There 259 of the carbon stars are classified as 
the \CR\ stars and 108 are \cn\ stars. We suggest to use F4600/F8000 the median flux ratio of the blue (4300$-$4600\,\AA) and red (8000$-$8700\,\AA) to 
select the \cn\ stars with completeness of 100\% and contamination rate of only 16\%. We demonstrate that although the 2MASS photometry 
is able to identify \cn\ stars with reasonably low contamination, it may 
have relatively lower completeness.

It is very interesting to do follow-up time-domain photometric and high-resolution spectroscopic observations in the future 
in order to identify carbon stars and further investigate their nature.

\acknowledgments

We thank the anonymous referee for positive and constructive comments which helped to improve this paper greatly, and thank Dr. Sarah Bird for her kind help in language editing. This work was supported by the National Key Basic Research Program of China 2014CB84570, the National Natural Science Foundation of China (NSFC) under grant U1231119, 11321064, 11390371, 11273011, 11473033,
and the China Postdoctoral Science Foundation under grant 2013M531587. CL acknowledges the Strategic Priority Research 
Program ``The Emergence of Cosmological Structures" of the Chinese Academy of Sciences, Grant No. XDB09000000 and the NSFC under grants 11373032 and 11333003.
The Guoshoujing Telescope (the Large Sky Area Multi-Object Fiber Spectroscopic Telescope LAMOST) is a National Major Scientific Project built by the Chinese Academy of Sciences. Funding for the project has been provided by the National Development and Reform Commission. LAMOST is operated and managed by the National Astronomical Observatories, Chinese Academy of Sciences. This research has made use of the SIMBAD database, operated at CDS, Strasbourg, France.

\clearpage

\begin{table*}
	\begin{center}
		\caption{Line indices definition used in this work}\label{tab:lineindices}
		\begin{tabular}{c|c|c}
			\hline\hline
			Name & Index Bandpass ($\rm\AA$) & Pseudo-continua ($\rm\AA$)\\
			\hline
			Ba II & $4550-4559$& $4538-4545$ $4559-4563$\\
			\ctwo\ & $5235-5640$ & $5165-5235$ $5640-5730$\\
			CN7065 & $7065-7190$ & $7025-7065$ $7190-7230$\\
			CN7820 & $7820-8000$ & $7790-7820$ $8000-8065$\\
			\hline\hline
		\end{tabular}
	\end{center}
\end{table*}
\clearpage

\begin{table*}[htbp]
	\centering
	\caption{Steps to identify carbon stars.}\label{tab:steps}
	\begin{tabular}{l|c|cc}
		\hline\hline
	    Step & Criteria &  \multicolumn{2}{c}{Selected candidates} \\
		\hline
		1 & S/N$(i)>10$ and leave only one epoch for multiply observed stars &  \multicolumn{2}{c}{2\,434\,289} \\
		\hline		
		2 & Equations (2), (3), (4) &  \multicolumn{2}{c}{87\,160} \\
		\hline
		\multirow{2}{*}{3}& CN7065\,$\ge4$ or CN7820\,$\ge8$  & 4\,490  &\multirow{2}{*}{27\,789}  \\
		\cline{2-3}
		& CN7065\,$<4$ and CN7820\,$<8$ with S/N$(g)>20$ and S/N$(i)>20$   & 23\,299 \\
		\hline
		4 &  CN7065\,$\ge2$ and \ctwo\,$\ge-13$ & \multicolumn{2}{c}{11\,254}  \\
		\hline
		5 & $K_s<14.5$ and $J-K_s>0.45$ &  \multicolumn{2}{c}{6\,309}  \\
		\hline
		6 & inspect by eye &  \multicolumn{2}{c}{894} \\
		\hline\hline
	\end{tabular}
\end{table*}

\clearpage

\begin{table*} 
	\begin{center}
		\caption{Carbon stars discovered in LAMOST DR2}\label{tab:carbonstars}
		\resizebox{\textwidth}{!}{ 
		\begin{tabular}{c|c|c|c|c|c|c|c|c|c|c}
			\hline\hline
			Obsid & Designation &RA & Dec & $J$ & $H$ & $K_s$ & SpType\tablenotemark{a} & SIMBAD & H$_\mu$& Giant/Dwarf\\
			\hline
1605200  & J003858.70+394504.1  & 9.7446096  & 39.751164  & 12.816  & 12.292  & 12.218  & C-H  &   & 0.874  &  \\
8104129  & J093547.66+270939.2  & 143.948584  & 27.160913  & 14.041  & 13.533  & 13.385  & C-H  & C*\tablenotemark{b}  & 12.554  & Dwarf\\
8106104  & J094602.02+265255.8  & 146.508422  & 26.882177  & 14.318  & 13.843  & 13.717  & C-H  & C-H  & 2.975  &   \\
14902109  & J004619.17+354537.1  & 11.579886  & 35.760306  & 11.795  & 11.252  & 11.115  & C-H  &   & 1.099  &   \\
2508139  & J045413.85+274714.0  & 73.557731  & 27.787226  & 12.334  & 11.564  & 11.277  & C-N  &   & 4.622  &   \\
3112230  & J060024.07+311152.5  & 90.100308  & 31.197925  & 8.592  & 7.408  & 6.924  & C-N  &   & 0.285  &   \\
8505213  & J054646.74+261235.6  & 86.694754  & 26.209909  & 13.52  & 12.408  & 12.02  & C-N  &   & 3.057  &   \\
1614232  & J003251.55+410906.9  & 8.2148293  & 41.15194  & 12.157  & 11.724  & 11.636  & C-R  &   & 6.311  & Dwarf\\
2615079  & J064054.96+300432.4  & 100.229  & 30.075667  & 12.15  & 11.485  & 11.305  & C-R  &   & 4.272  &   \\
2616122  & J064149.09+303309.8  & 100.45456  & 30.552731  & 11.427  & 10.754  & 10.575  & C-R  &   & 3.147  &   \\
236809201  & J132315.71+034347.3  & 200.81548  & 3.729823  & 7.59  & 7.054  & 6.866  & C?\tablenotemark{c}  & C-R  & 3.456  &   \\
238005083  & J184953.52+453155.0  & 282.473025  & 45.53195  & 9.764  & 9.229  & 9.1  & C?  &   & 3.157  &   \\
240806239  & J163859.12+093401.2  & 249.746369  & 9.567  & 11.248  & 10.676  & 10.574  & C?  &   & 2.846  &   \\
28006203  & J045019.27+394758.7  & 72.580329  & 39.799648  & 12.193  & 11.459  & 11.261  & Unknown\tablenotemark{d}  &   & 3.055  &   \\
30116006  & J064624.52+075621.9  & 101.6022  & 7.939422  & 12.023  & 11.415  & 11.282  & Unknown  &   & 2.035  &   \\
...& ...&... & ... &... &... & ...&...  & ...& ...&...\\
 			\hline\hline
		\end{tabular}}
	\end{center}
	\tablenotetext{a}{Spectral sub-type}	
	\tablenotetext{b}{Carbon stars marked in SIMBAD with no sub-types identified}
	\tablenotetext{c}{Suspect carbon stars}
\tablenotetext{d}{Carbon stars with no identifiable sub-type}
(This table is available in its entirety in a machine-readable form in the online journal. A portion is shown here for guidance regarding its form and content.)

\end{table*}

\begin{table}
	\centering
	\caption{Criteria for classifying carbon stars.}\label{tab3}
	\begin{tabular}{c|l}
		\hline\hline
		Sub-type & Criteria\\
		\hline
		\multirow{3}{*}{C-N} & 1) No flux at $\lambda<4400$\,$\rm\AA$; some very late type C-N can be flat even at $\lambda<5000$\,$\rm\AA$;\\
		& 2) Strong Ba II at 6496\,$\rm\AA$;\\
		& 3) Weak H$\alpha$.\\
		\hline
		\multirow{4}{*}{C-H} & 1) Strong G-band (CH); \\
		& 2) Strong CN at 4215\,$\rm\AA$ and weak Ca at 4227\,$\rm\AA$;\\
		& 3) Strong Ba II at 4554\,$\rm\AA$ or 6496\,$\rm\AA$;\\
		& 4) Strong H$\alpha$.\\
		\hline
		\multirow{2}{*}{C-R} & 1) Strong CN at 4215\,$\rm\AA$ and strong Ca at 4227\,$\rm\AA$;\\
		& 2) Weak Ba II at 4554\,$\rm\AA$; relatively weak Ba II at 6496\,$\rm\AA$ compared to H$\alpha$.\\
		\hline\hline
	\end{tabular}
\end{table}

\clearpage

\clearpage

\clearpage

\clearpage

\end{document}